\documentclass[review]{elsarticle}

\usepackage{pdfpages}
\usepackage[utf8]{inputenc} 
\usepackage{graphicx} 
\usepackage{makeidx} 
\usepackage{xcolor} 
\journal{**********}

\begin{document}

\begin{frontmatter}

\title{Unidirectional reflectivity in non-Hermitian lossy thin films}

	\author[mymainaddress,mysecondaryaddress]{A. Padr\'on-God\'inez}
	\ead{apadron@inaoep.mx, alpago00@unam.mx}
	\author[mymainaddress]{C. G. Trevi\~no-Palacios}
	\cortext[cor1]{Corresponding author.}
	\address[mymainaddress]{Instituto Nacional de Astrof\'isica, \'Optica y Electr\'onica, Luis Enrique Erro No. 1,
	Santa Mar\'ia Tonantzintla, C.P. 72840, Puebla - M\'exico 
}
	\address[mysecondaryaddress]{Instituto de Ciencias Aplicadas y Tecnología, 
	Universidad Nacional Aut\'onoma de México, Circuito Exterior S/N, Coyoacán, A. P. 04510, Cd. Universitaria - M\'exico
}	




\begin{abstract}
Inspired by non-Hermitian systems, we study reflection and transmission in a stack of thin films composed by the repetition of a bipartite unit cell. We aim for controlled reflection and transmission using lossless and lossy materials in order to develop an optical diode to generate entanglement photons source. Particularly, we show unidirectional reflection using transfer matrix methods and confirm our results by finite element simulation. 
\end{abstract}

\begin{keyword}
\texttt{Controlled reflection, thin films, non-Hermitian optics system}
\end{keyword}

\end{frontmatter}


\section{Introduction}
Thin films are a standard option to design optical devices with controlled reflection and transmission. Optics has a long tradition of studying how the basic properties of materials can be used to engineer thin film structures with an overall different behavior \cite{Knittl}, which might be used for industrial applications such as optical camouflage and optical rectifiers, isolators or switches\cite{Feng, Ramezani} to mention a few.  
Recently, the quantum idea of PT-symetry has been used to create composite structures with interesting optical properties. PT-Symmetry in quantum mechanics refers to invariance to spatial and temporal reflection. This is provided by complex potentials that obey the property V*(x)=V(-x) \cite{Bender, Lin}. An ideal optical equivalence is to introduce linear media with equal real part of the refractive index and imaginary parts that are the complex conjugate of each other \cite{Kottos}. Such media is practically inexistent in nature and hard to engineer but experimental realizations have shown its feasibility\cite{Feng}.
Furthermore, it has been shown that unidirectional reflection arises from PT-symmetric structures due to the gain-loss balance in optical structures that bring to mind a stack of thin film \cite{Longhi}.

\begin{figure}
	\centering 
	\includegraphics[height=5cm,width=10cm ]{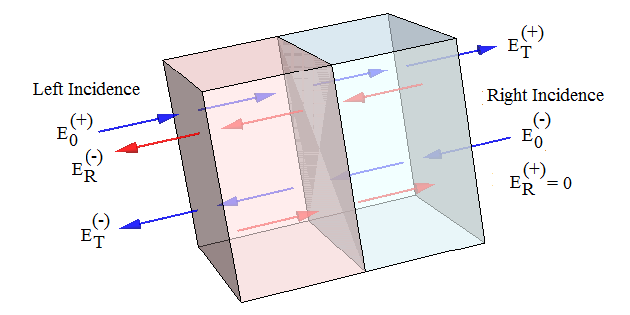}
	\caption{A unit cell showing unidirectional reflectivity using two thin-films with conjugate.}\label{fig1_uni}
\end{figure}

Here, we are interested in the effect of using real-world materials to design unidirectional reflectionless stacks of thin films, Fig.(\ref{fig1_uni}).
First, we are going to model electromagnetic field propagation through dielectric thin films considering left and right normal incidences.
We will start with the treatment of a simple layer using multiple beam interference techniques.
Then we will provide their transfer matrixs  and use them to describe a unit cell composed of just two thin layers \cite{Steck}.
Next, we will optimize the film thicknesses numerically to find the extremal values for reflectivity/transmitivity at the desired wavelength.
Then, we will find the transfer matrix results for these optimized parameter values and compare them with finite element simulation.
We will use the ideal gain-loss bilayer as benchmark for more realistic passive-loss values of doped silicon dioxide ($SiO_2$) with metal nanoparticles for loss and erbium ions for gain.
Finally, we will show results for unidirectional reflectance for a stack composed by three unit cells. 
\section{Transfer matrix formalism}
\begin{figure}
	\centering 
	\includegraphics[height=5cm,width=9.5cm]{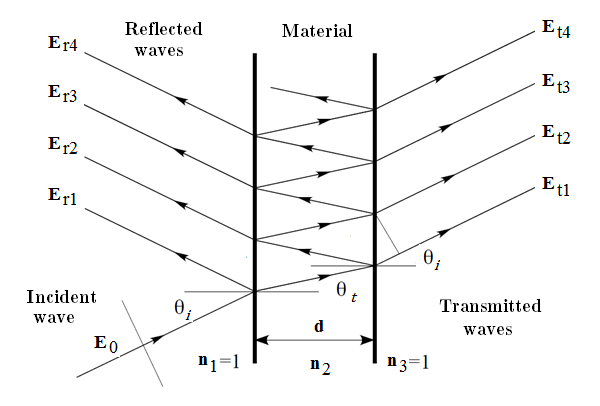}
	\caption{Successive reflections and transmissions in a thin film with an incidence angle.}\label{fig2_inci}
\end{figure}
We start our study with the treatment of a single thin layer using multi-beam interferometry \cite{Ramezani}. We consider incidence from left to right, Fig.(\ref{fig2_inci}), and obtain:
\begin{eqnarray}
t_{13} &=&
\frac {  t_{23}t_{12}e^{ i \phi}}{1-r_{23} r_{21}e^{i \phi}}\label{eq:tab},
\nonumber\\
r_{13} &=&
r_{12}+ \frac {  t_{21} r_{23}t_{12}e^{ i \phi}}{1-r_{23} r_{21}e^{i \phi}}\label{eq:rab},
\end{eqnarray}
where $t_{ij}$ and $r_{ij}$, are the transmission and reflection coefficients at each boundary, with $i \neq j$.
We will show results for normal incidence for the sake of space. Here the effective transfer matrix \cite{Longhi}, also called scattering matrix is given by 
\begin{eqnarray}
S_{e}={\left( \begin{array}{cc}\frac{ t_{12} t_{23}e^{i  k_2 {n_2} {d_2}}}{1-r_{21}r_{23}e^{2 i  k_2 {n_2} {d_2}}} &r_{21}+ \frac {  t_{12} r_{32}t_{21}e^{i  k_2 {n_2} {d_2}}}{1-r_{23} r_{21}e^{i  k_2 {n_2} {d_2}}} \\ r_{12}+\frac {  t_{12} t_{23}r_{23}e^{i  k_2 {n_2} {d_2}}}{1- r_{21} r_{23}e^{ i  k_2 {n_2} {d_2}}} &  \frac {  t_{32}t_{21}e^{i  k_2 {n_2} {d_2}}}{1-r_{23} r_{21}e^{i  k_2 {n_2} {d_2}}} \end{array}\right)}=\left(\begin{array}{cc} t_{13} & r_{31} \\ r_{13} & t_{31}\end{array}
\right)
\end{eqnarray}\\
where $k_2$ is the wavenumber in the material, $d_2$ is the width of the thin layer, $n_2$ is the refraction index, $t_{13}$ and $r_{13}$ are the transmission and reflection coefficients for left side incidence then $r_{31}$ and $t_{31}$ are the reflection and transmission coefficients considering right side incidence.
\begin{figure}
	\centering 
	\includegraphics[height=3.0cm,width=12cm]{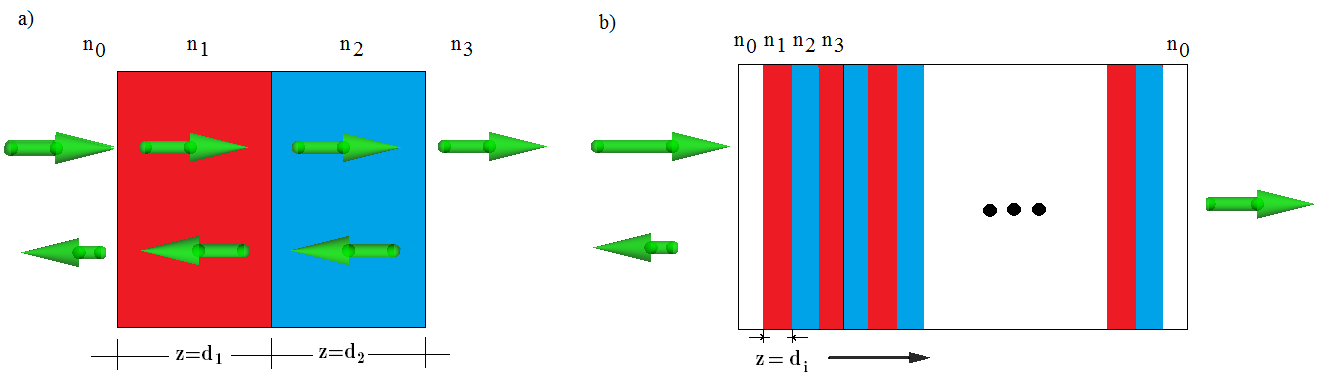}
	\caption{a) Unit cell N=1.  b) Stack of unit cells N=1,2,3,... }\label{fig3_dim}
\end{figure}
When we have a layer with complex refractive index, $n_{ci}$ =$ n_i \pm i\cdot\kappa_{i}$, a negative imaginary part, $\kappa_{i}<0$, provides gain and loss is obtained with $\kappa_{i}>0$. In addition, transmission and reflection for a field impinging on the right side are given by $t_{13}$ and $r_{13}$ and for a left impinging field $r_{31}$ and $t_{31}$.

\section{Semiconductor and Thin Films}
\begin{figure}
	\centering 
	\includegraphics[height=2.75cm,width=9.5cm]{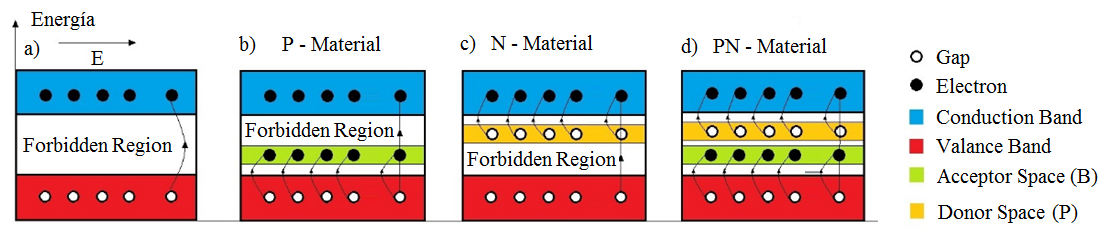}
	\caption{The conduction and valence bands of pure Silicon.}\label{fig_4}
\end{figure}

The basic material used in the construction of most diodes and transistors is Silicon (Si), silicon is a semiconductor at room temperature very few electrons exist in the conduction band of the silicon crystal. When a proportional current is applied to a group of moving electrons, the current is small, the material has great resistance. The conduction and valence bands of pure silicon are shown in figure (\ref{fig_4}).
At 0[K] (absolute zero), all electrons are at their lowest energy level, at room temperature occasionally an electron has a lot of energy to escape the valence band and move towards the conduction band. The lack of electrons is shown as a circle or hole, if an electric field is applied to the material the electron moves towards the positive terminal of the battery. An electron in the valence band can also move towards the positive terminal of the battery if it has enough energy to go from its energy level to the energy level of the hole.
When this electron escapes from a gap, it leaves it hidden. The gaps would appear to be moving to the right, towards the negative battery terminal. The current network is therefore the sum of the current due to the movement of the electrons in the conduction band and of the current due to the movement of the gaps in the valence band [14]. The conventional current due to the flow of electrons and the current of gaps are in the direction of the electric field.

\subsection{Semiconductor and Thin Films}
Similarly, the use and procedure we make for the construction of an electric diode is used to create thin films that will serve to create an optical diode in this case and to carry out on this device the study and analysis when electromagnetic radiation is applied by two normal directions opposite. In thin films we look first to have symmetry-PT, after doping with other materials. Which we accomplish with a pair of semiconductor dielectric (dimer) layers with balanced gain and loss. However, to achieve a symmetry of the device we must optimize its thickness using parameters on which we want to obtain the results, such as wavelength, complex refractive indices, to name a few. Using thin film layers, we can use various combinations of passive-doping as gain-gain in the dimer, or loss-loss, up to layers such as gain-loss, real-loss, real-gain \cite{Poladian}.

\subsection{Design and construction of study materials}
For the analysis of this work, we look for materials based on semiconducting dielectric thin films, which have a behavior similar to that of a signal rectifier diode. The basis of these materials were originally thin layers of Silicon and Germanium. The material studied starts from $SiO_2$ that has the following properties when thin films are deposited on a substrate with refractive index $n_r = 1.4574$, an extinction coefficient $\kappa = 0.000687$, at a wavelength of 1510 [nm], \cite{Rodriguez}. This material is constructed by means of a reactive electron beam, evaporated on various substrates at 300 [$^oC$], as is done in the manufacture of CMOS transistors \cite{Hodgson}. We build a first material similar to that shown in figure (4 d), which has an acceptor and donor in the valence bands as a rectifier diode that is polarized in direct current  \cite{Padron3}. Generating a PN-material with Boron and Phosphorus, which corresponds to a material with balanced gain and loss based on a pair of thin films, known as a dimer with parity time symmetry (PT).
We also require a second material constructed with acceptor valence bands like the one in figure (4b), doped with Boron as an N-material. Thus creating a pure material in its first layer and with absorption in the second layer, which corresponds to a dimer with a passive layer and a layer with loss. These materials that have been tested at the electrical system level, we now study them to create optical rectifiers that are important in the construction of entangled photon generating sources. The properties of these dimers based on doped SiO2 dielectric thin films have dielectric constants equal to
\begin{eqnarray}
\varepsilon =\varepsilon _0 \pm i \varepsilon _i
\label{eq:rabis},
\end{eqnarray}
where $\varepsilon _0 = 2.3963$ and $\varepsilon_i = 0.3003$. So we will use for a balanced dimer $\varepsilon =\varepsilon _0 - \left(i \varepsilon _i\right)$ in the first layer and $\varepsilon =\varepsilon _0 + \left(i \varepsilon _i\right)$ in the second layer, gain-loss. Then for a passive dimer with loss we have $\varepsilon =\varepsilon _0$ in the first layer and $\varepsilon =\varepsilon _0 + \left(i \varepsilon _i\right)$ for the second layer, real-loss \cite{Shramkova}. Then we change these values to refractive index and extinction coefficient respectively ($n_i=1.548$, $\kappa_i=0.548$).

\section{Optimization}
\begin{figure}
	\centering 
	\includegraphics[height=5.0cm,width=10cm ]{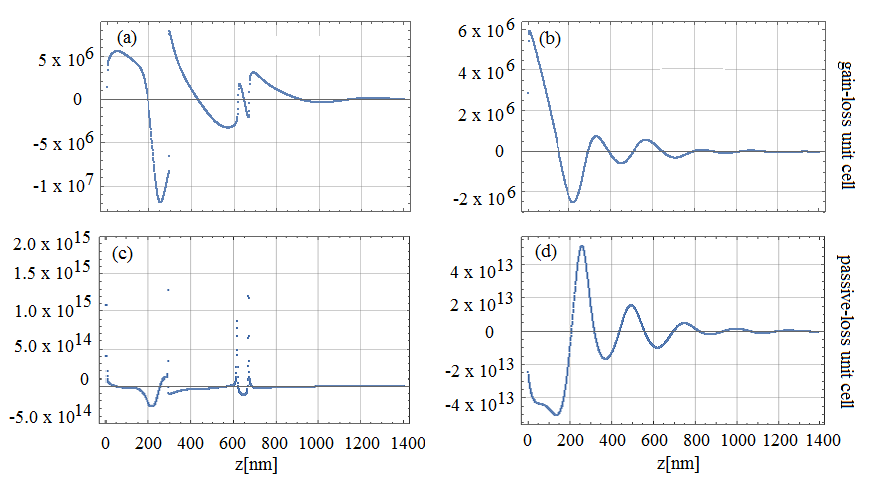}
	\caption{Normal incidence reflection coefficient (R20) (a) first numerical derivative, and (b) second  numerical derivate for  one gain-loss unit cell, (c) First numerical derivative, and (d) second  numerical derivate for  one passive-loss unit cell. Notice the narrowing in the shape on the gain-loss (upper) system in comparison with the passive-loss system (lower).}\label{fig4_dn1}
\end{figure}

We are looking for unidirectional reflectionless \cite{Shramkova,Yang}, that is a reflection that is null in one direction but not in the other. For fixed refractive indexes and impinging wavelength, we can optimize the dimer thickness for reflection or transmission using the first and second numerical derivatives of their transfer matrixs. Figure (\ref{fig4_dn1}) shows numerical derivatives for the reflection and transmission coefficients of a double-layer cell with balanced gain-loss and one double-layer cell with passive-loss materials at $\lambda=1550$ [nm] and normal incidence, $\theta_{i}=0$ .
The parameters used for the ideal cell were $n_1=n_2=1.548$, of silicon dioxide ($SiO_2$) doped with either metal nanoparticles for loss ($\kappa_{2}$) or erbium ions for gain ($\kappa_{1}$) with  $\kappa_{1}=-\kappa_{2}=-0.548$. For the passive-loss cell $n_1=n_2=1.548$, $\kappa_{1}=0$, $\kappa_{2}=0.548$, in both cases $n_0=n_3=1$, as we showed Fig.(\ref{fig3_dim}), \cite{Novitsky}. The thickness of the layers are equal, $d_1=d_2=d$. 
We take the zeros of the first derivative that yield positive values of second derivative to find the layer thickness that yields reflectivity minima.
When the electromagnetic waves propagate within a dielectric medium, the phases $\phi$ are cumulative and depend on the refractive index $n_{ci}$, the wavenumber $k_0$ and the width of the layer that in turn it depends on the wavelength $d_i(\lambda)$. For example, the optimized thickness for reflectivity minima for the balanced bilayer are d=220[nm] and d=660[nm] at the desired wavelength of 1550[nm]. The optimized thickness for transmission minima for passive-lossy cell are the d=146[nm] and d=200[nm] at the same desired wavelength.  
In the following we will simulate the propagation of linearly polarized electromagnetic field using these optimized thickness and compare our numerical results with finite element simulation to good agreement.

\section{Results}
The system with which we start, as in equation (3) for the case of two layers is the transfer matrix now given by
$M_{e n t}=M_{23}\cdot M_{22}\cdot M_{12}\cdot M_{11}\cdot M_1$
with it we get the effective dispersion matrix, in terms of the reflection and transmission coefficients\\

\begin{eqnarray}
S_e=\frac{1}{m_{22}}\left(
\begin{array}{c}
Det \left(M_{net}\right) \hspace{.3cm} m_{12} \\
-m_{21}        \hspace{.9cm}  1\\
\end{array}
\right)
=\left(
\begin{array}{c}
t_{03} \hspace{.3cm}  r_{30} \\
r_{03} \hspace{.3cm}  t_{30} \\
\end{array}
\right)
\end{eqnarray}

that we can express it as the system of equations to analyze the propagation in the direction of the z-axis to a quarter of the resonance wavelength

\begin{eqnarray}
\left(\begin{array}{c} E_{3}^{(+)}  \\ E_{0}^{(-)}\end{array}\right)
=\frac{\lambda_r}{4 n_e }S_e
\left(\begin{array}{c} E_{0}^{(+)}  \\ E_{3}^{(-)}\end{array}\right)
\end{eqnarray}

From the effective scattering matrix, the analysis of the eigenvalues is performed for one quarter of the wavelength propagation. This analysis is similar to the autonomous treatment of a system of ordinary second-order differential equations of the form
\begin{eqnarray}
\frac{dE_3^{+}}{dz}=f(r,t) \hspace{0.5cm}
\frac{dE_0^{-}}{dz}=g(r,t)
\end{eqnarray}

For the roll that is being analyzed, the effective scattering dispersion matrix contains reflectivity and transmitivity based on complex refractive indexes. As presented the loss or gain extinction coefficient $\kappa_i$ respectively is the parameter that can balance the optical device based on thin dielectric films for photon entanglement source. Therefore, the trajectories of the eigenvalues of the effective scattering matrix are presented below.\\
To illustrate the PT-Symmetry in the figure ($6b$ and $6e$) the eigenvalues $\lambda_1$ y $\lambda_2$ with reals and imaginary parts, show reals values of 1 $\leq$ ($\lambda_1$, $\lambda_2$) from $\kappa_2=-0.548$ and 0 $\leq$ ($\lambda_1$, $\lambda_2$) in the range $-0.548 \leq \kappa_2 \leq 0.548$ in the imaginary part of ($\lambda_1$, $\lambda_2$).

\begin{figure}
	\centering 
	\includegraphics[height=6cm,width=12.5cm]{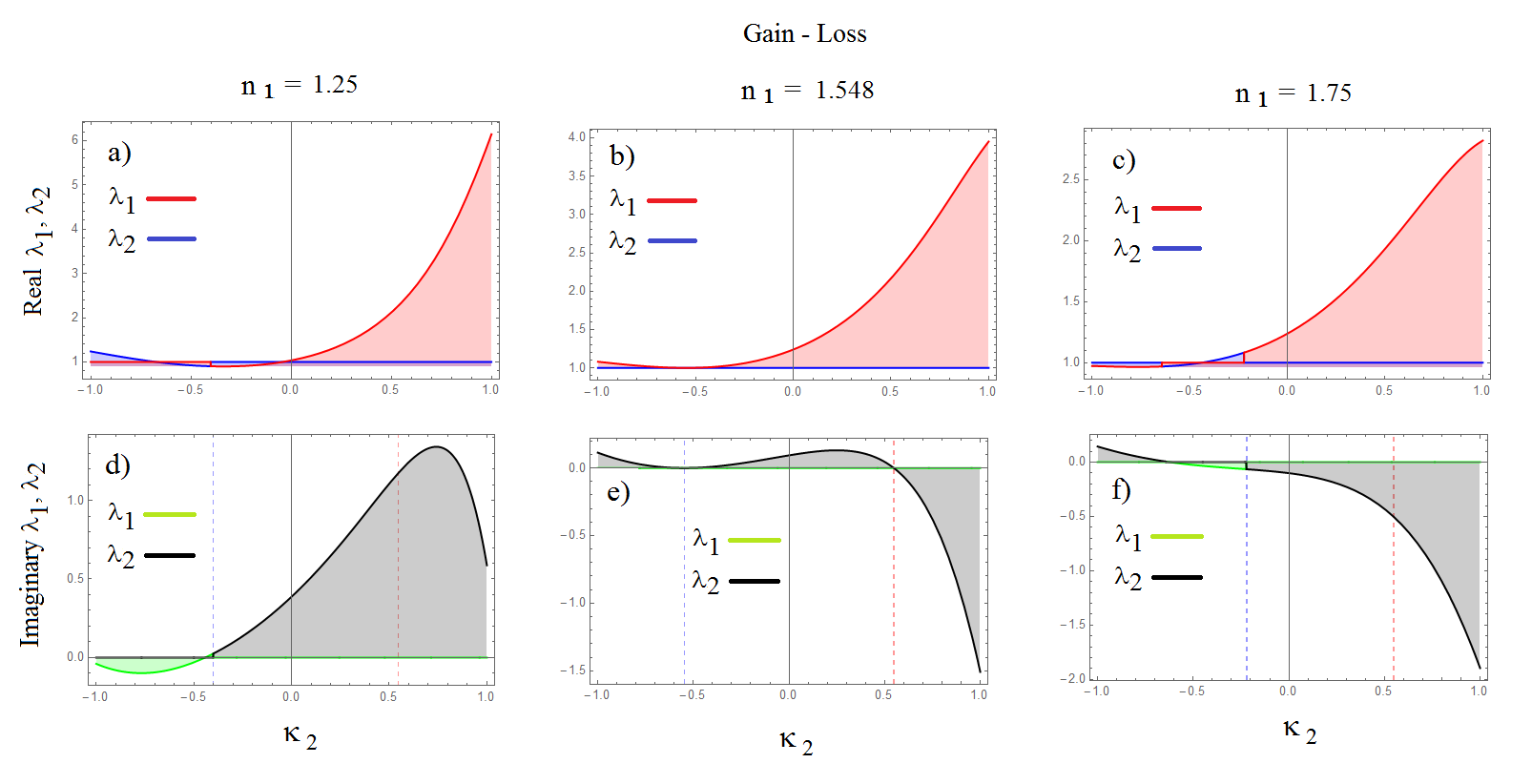}
	\caption{Eigenvalues $\lambda_1$, $\lambda_2$ versus extinction coefficient $\kappa_2$ gain-loss, reals a), b) and c); imaginaries d), e) y f). }\label{fig_6}
\end{figure}

\begin{figure}
	\centering 
	\includegraphics[height=6cm,width=12.5cm]{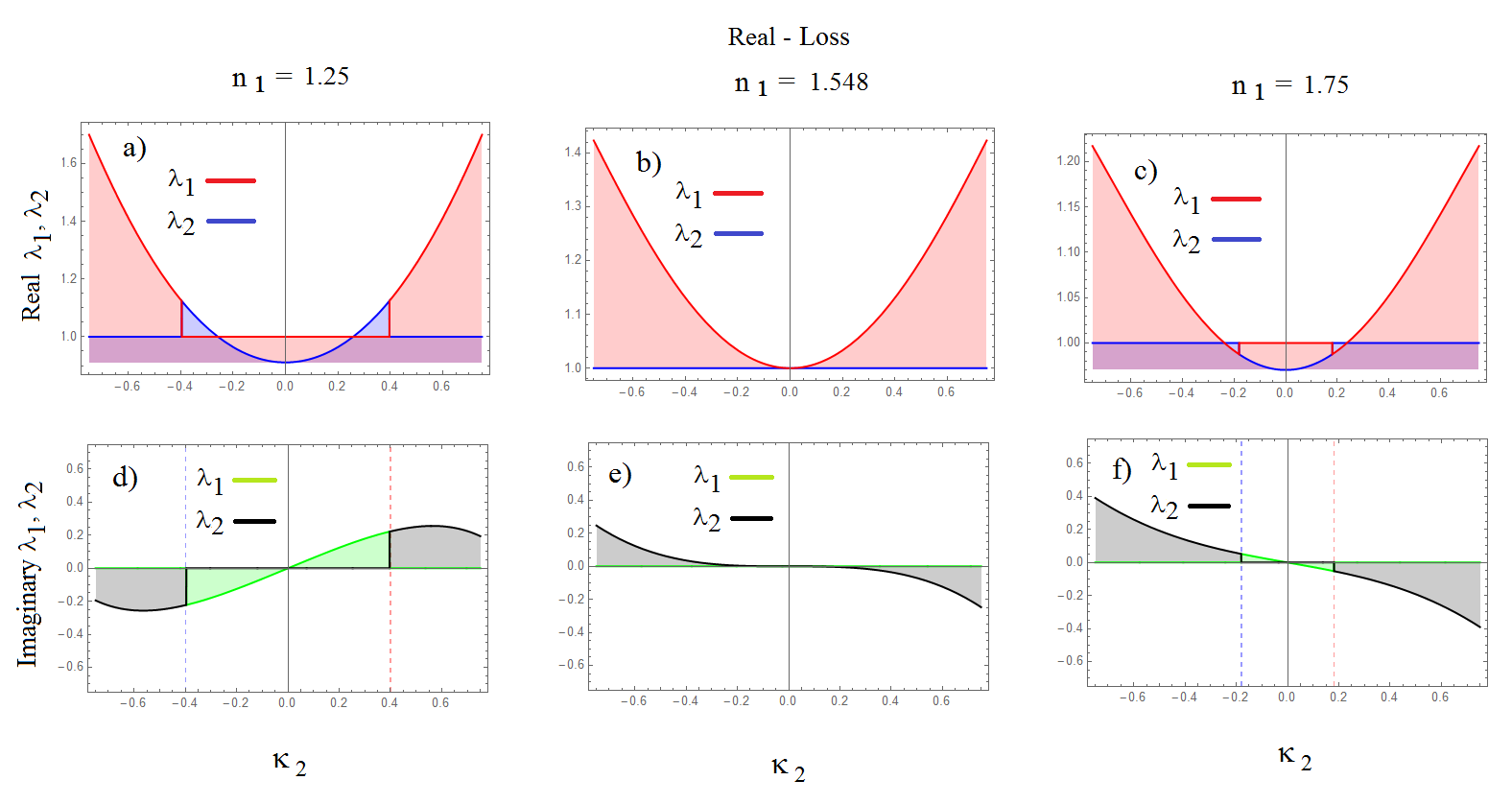}
	\caption{Eigenvalues $\lambda_1$, $\lambda_2$ versus extinction coefficient  $\kappa_2$ gain-loss, reals a), b) y c); imaginaries d), e) y f).}\label{fig_7}
\end{figure}

\begin{figure}
	\centering 
	\includegraphics[height=5cm,width=12.0cm]{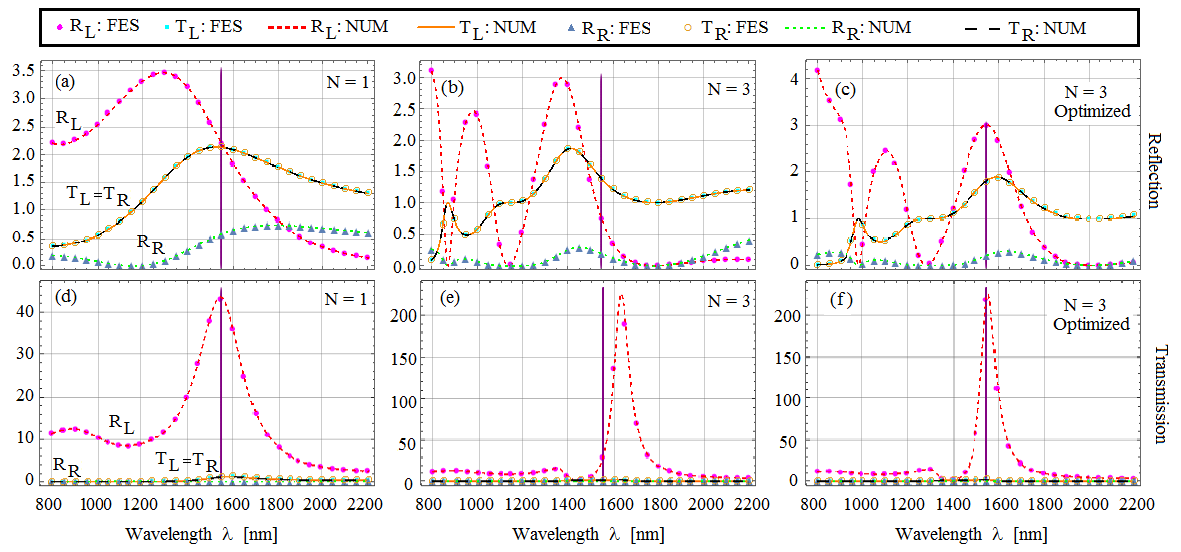}
	\caption{ Numerical (NUM) and finite element simulation (FES) for counterpropagating reflectivity and transmitivity in an ideal balanced double-layer cell. Calculations for one bilayer (N=1) with (a) d=220 [nm] and (d) d=660 [nm], and 3 bilayers with and (b) d=220 [nm] and (e) d=660 [nm]. An optimized 3 bilayers system is obtained with (c) d=248 [nm] and (f) d=630 [nm]. Notice the changes in left reflectivity ({$R_L$}) in comparison with right reflectivity ({$R_R$}) and that the transmissions are equal ({$T_L$}={$T_R$}). The vertical line represents the target 1550 [nm] wavelength.}\label{fig5_cpi1}
\end{figure}

Figure (\ref{fig5_cpi1}) shows the reflectivity and transmitivity coefficients for ideal gain-loss structures. Figure  (\ref{fig5_cpi1}a) shows these values for an optimized bilayer thickness of $d=220[nm]$ where we can see that reflectivity from the left-side, $R_L$, dominates over that from the right-side, $R_R$, at $\lambda=1550[nm]$, which is showed as a vertical line. If we wanted to use this bilayer and repeat it as unit cell in a stack, say $N=3$, the wavelength showing a maximum difference in reflection will shift, Fig. (\ref{fig5_cpi1}b). Thus, we have to optimize for each and every stack size to recover similar results to the $N=1$ case, Fig. (\ref{fig5_cpi1}c) with $d=248[nm]$.
Similarly, Fig.  (\ref{fig5_cpi1}d) shows reflectivity and transmitivity values for an optimized bilayer thickness of $d=660[nm]$ where we can again see that reflectivity from the left-side dominates over that from the right-side. Again, a wavelength shift occurs if we increase the number of unit cells without further optimization,  Fig. (\ref{fig5_cpi1}e) and when we optimized for N=3 with $d=630[nm]$ we recover results as case N=1,  Fig. (\ref{fig5_cpi1}f).

\begin{figure}
	\centering 
	\includegraphics[height=5cm,width=12.0cm]{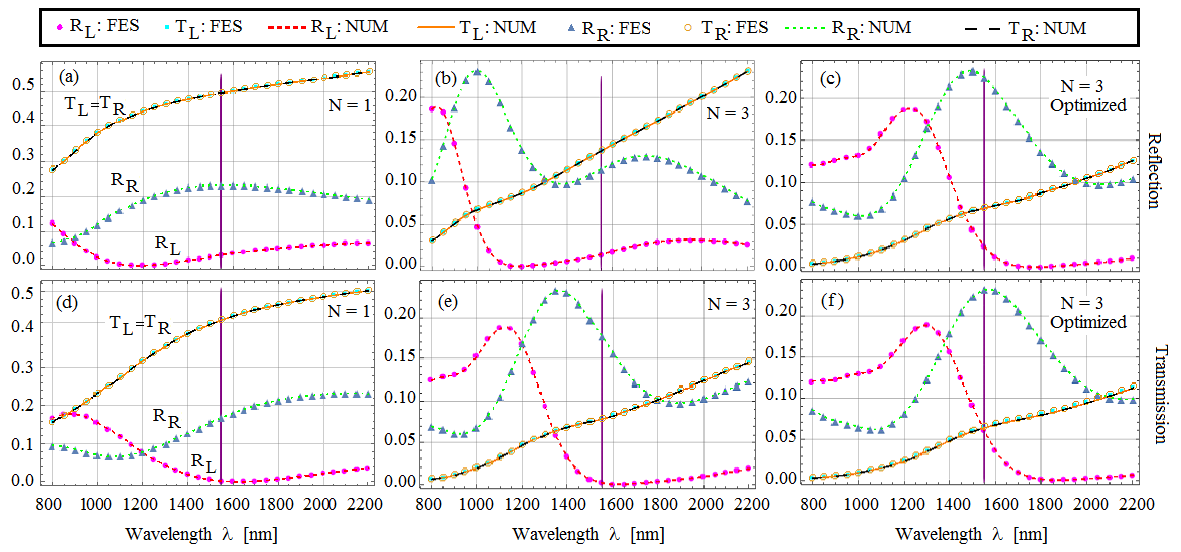}
	\caption{ Numerical (NUM) and finite element simulation (FES) for counterpropagating reflectivity and transmitivity in an ideal balanced double-layer cell. Calculations for one bilayer (N=1) with (a) d=146 [nm] and (d) d=200 [nm], and 3 bilayers with and (b) d=146 [nm] and (e) d=200 [nm]. An optimized 3 bilayers system is obtained with (c) d=218 [nm] and (f) d=230 [nm]. Notice clearly that the left reflectivity is bigger than right reflectivity, (${R_L}>{R_R}$) and moreover the transmissions ({$T_L$}={$T_R$}). The vertical line represents the target 1550 [nm] wavelength.} \label{fig6_cpr1}
\end{figure}

Now, we move into a more realistic passive-loss structure. Figure (\ref{fig6_cpr1}a) shows the reflectivity and trasmitivity values for an optimized bilayer thickness of $d=146[nm]$ where we can see now that,  reflectivity from the right dominates over that from the left at $\lambda=1550[nm]$. An equivalent wavelength shift is induced when we stack the unit cell N=3 without further optimization, Fig.(\ref{fig6_cpr1}b), but this can be easily corrected with optimization for the new stack, Fig.(\ref{fig6_cpr1}c)  with $d=218[nm]$.
Furthermore, Fig.(\ref{fig6_cpr1}d) shows the reflectivity and transmitivity coefficients for passive-loss structures. There we show these values for an optimized bilayer thickness of $d=220[nm]$ where we can see that the reflectivity $R_R$, dominates over $R_L$ at $\lambda=1550[nm]$ (vertical line). If we wanted to use this bilayer and repeat it as unit cell in a stack, say $N=3$, the wavelength showing a difference in reflection will shift, Fig.(\ref{fig6_cpr1}e). Thus, we have to optimize for each and every stack size to recover similar results to the $N=1$ case, Fig.(\ref{fig6_cpr1}f) when $d=230[nm]$.
\section{Conclusions}
We have shown that optimization of thin film structures, using transfer matrix analysis, can yield optimal parameters for structures with bidirectional transmission but unidirectional reflection, as an optical diode (mirror of one face). We presented two cases for two different optimal values of thin film thickness. One being the bipartite unit cell with balanced gain-loss as a one dimer with PT-symetric and the other a passive-loss structure. Finally, we want to stress that the width of the unit cell must be optimized for the desired stack size, otherwise unidirectional reflectivity will occur at a different wavelength. 
We desired the resonnance at $\lambda=1550[nm]$ for optical communications.
Of course, in the more realistic passive-loss unit cell, the transmittance will decrease with the size of the stack as a result of the losses. In addition for  both circunstances of unit cell, we noted the differences between unidirectional reflectivity; one case by the left side (dimer: balanced) and other by the right side (dimer: passive-loss).

\section*{Acknowledgements}
This work has been sponsored by “Dirección General de Asuntos del Personal Académico de la Universidad Nacional Autónoma de México” under the “Programa de Apoyos para la Superación del Personal Académico” (PASPA) through doctoral scholarship. The author thanks B. M. Rodríguez-Lara for fruitful discussion.

\section*{References}

\end{document}